\documentclass{webofc}
\usepackage[dvipsnames]{xcolor}

\usepackage[varg]{txfonts}   
\usepackage{hyperref}
\usepackage{url}

\hypersetup{colorlinks=true,citecolor=black,urlcolor=black,linkcolor=black}

\begin{document}
\title{Searching for the QCD critical point through constant entropy density contours}
%
%

\author{\firstname{Hitansh} \lastname{Shah}\inst{1}\fnsep\thanks{\email{hshah10@uh.edu}} ,
        \firstname{Mauricio} \lastname{Hippert}\inst{2} ,
        \firstname{Jorge} \lastname{Noronha}\inst{3},
        \firstname{Claudia} \lastname{Ratti}\inst{1}, \and
        \firstname{Volodymyr} \lastname{Vovchenko}\inst{1}
}

\institute{ Department of Physics, University of Houston, Houston, TX 77204, USA 
\and
           Centro Brasileiro de Pesquisas Físicas, Rio de Janeiro, RJ, 22290-180, Brazil
\and
           Illinois Center for Advanced Studies of the Universe \& Department of Physics, University of Illinois at Urbana-Champaign, Urbana, IL 61801-3003, USA
          }

\abstract{
We propose a novel method to locate the QCD critical point by constructing an expansion along contours of constant entropy density. Applying two independent analysis of lattice QCD data at zero baryon chemical potential, we find a critical point at $T_c = 114 \pm 7$~MeV and $\mu_{B_c} = 602 \pm 62$~MeV for an expansion truncated at order $\mu_B^2$. This approach is consistent with recent lattice QCD results up to $\mu_B/T \leq 3$. A more precise determination of the required expansion coefficients from lattice simulations will be essential for reliably establishing the location of the QCD critical endpoint.
 }

\maketitle
\section{Introduction}
\label{intro}

Mapping the QCD phase diagram of strongly interacting matter is a fundamental problem in high-energy physics. 
In particular, the search for the conjectured QCD critical point (CP) is among the primary motivations for many experimental programs, in colliders such as RHIC and FAIR, that are focusing their attention towards baryon-rich matter. Theoretically, the CP location is challenging to obtain from first principles as direct lattice QCD simulations at non-vanishing net baryon density are hindered by the fermion sign problem. 
At zero baryon chemical potential, lattice QCD simulations indicate an analytic crossover from hadronic gas to deconfined quark gluon plasma \cite{Aoki:2006we}. 
Recently, several techniques were developed to predict the critical point at finite and real net baryon density, both based on model estimates and directly from lattice data at zero and imaginary chemical potential \cite{Clarke:2024padetwo, Basar:2024qcd, Hippert:2024bhe, Fu:2020frg, Gunkel:2021dse, Gao:2021frg}.

In this work, we aim to extrapolate lattice QCD data available at vanishing baryon chemical potential $\mu_B=0$, by following contours of constant entropy density towards finite $\mu_B$. We choose to look at these contours because the entropy becomes multivalued in the regime of a first-order phase transition. Based on the crossings of these entropy density contours, one can obtain the first-order line as well as the critical endpoint at finite chemical potential, if a phase transition is found. We thus obtain the QCD critical point location based on an expansion of entropy density contours to order $\mathcal{O}(\mu_B^2)$, with coefficients obtained from lattice QCD simulations.  We propagate lattice QCD errors to obtain confidence regions for the CEP location, using two different methods.

\section{Methodology}
To follow the contours of constant entropy density, we expand the function $T_s$ such that:
\begin{equation}
\label{eq:Ts}
    s\big[T_s(\mu_B; T_0),\mu_B\big) = s\big(T_0,\mu_B=0\big)\,
\end{equation}
\begin{equation}
\label{eq:TsO2}
T_s(\mu_B;T_0) \approx T_0 + \alpha_{2}(T_0) \frac{\mu_B^2}{2} + \mathcal{O}(\mu_B^4),
\end{equation}
where the expansion is truncated up to $\mu_B^2$. The coefficient of expansion $\alpha_2$ is calculated at $\mu_B = 0$ for a constant value of $s$ and is given as follows:
\begin{equation}
\label{eq:C2}
    \alpha_{2}(T_0) \big|_{\mu_B=0} = -\frac{2T_0 \chi_2^B(T_0) + T_0^2 \chi_2^{B'}(T_0)}{s'(T_0)}.
\end{equation}
Here $\chi_2^B = \left[ \frac{\partial^2 (p/T^4)}{\partial (\mu_B/T)^2} \right]_T$
is the second order baryon number susceptibility, normalized by $T^2$, and $s$ is the entropy density, while primes denote temperature derivatives. For the calculation of $\alpha_2$, the thermodynamic quantities are obtained from continuum extrapolated results from the Wuppertal Budapest Collaboration presented in Refs. \cite{Borsanyi:2021PRL, Borsanyi:2014PLB} for $\chi_2^B$ and $s$, respectively. 

To obtain the location of the CP, we exploit the fact that the CP is an inflection point, which means that $\partial^2s/\partial^2T$ diverges. Also, both at the CP as well as on the spinodals,  $\partial s/\partial T$ also diverges. When we substitute Eq.~\eqref{eq:Ts} in these two conditions, we obtain the pair of equations $\left(\frac{\partial T_s}{\partial T_0}\right)_{\mu_B} = 0$ and $\left(\frac{\partial^2 T_s}{\partial T_0^2}\right)_{\mu_B} = 0$.
Using the expansion in Eq.~\eqref{eq:TsO2}, we find that
\begin{align}
    \label{eq:spinodal}
\alpha_2''(T_{0,c}) = 0, &&
\mu_{B,c} = \sqrt{-\frac{2}{\alpha_2'(T_{0,c})}}, && T_c = T_{0,c} + \alpha_2(T_{0,c}) \frac{\mu_{B,c}^2}{2}.
\end{align}
 From these conditions, we can obtain the critical point location in the $T-\mu_B$ plane.

\section{Results}

\begin{figure*}
    \centering
    \includegraphics[width=6cm,clip]{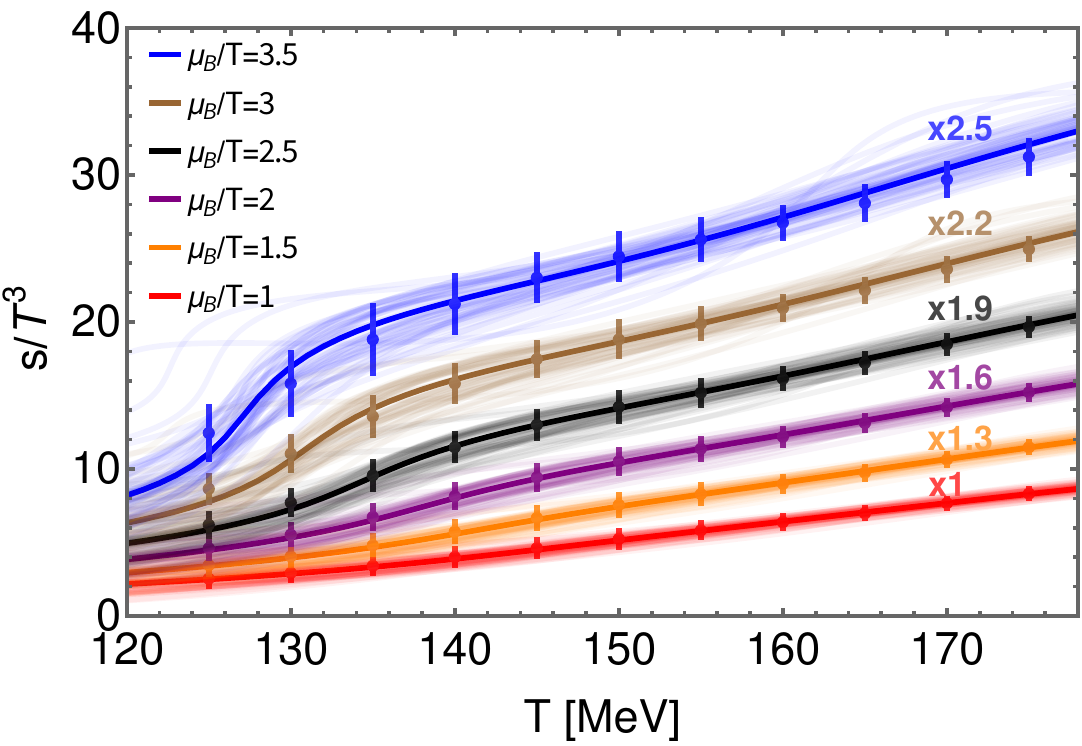}
    \includegraphics[width=6cm,clip]{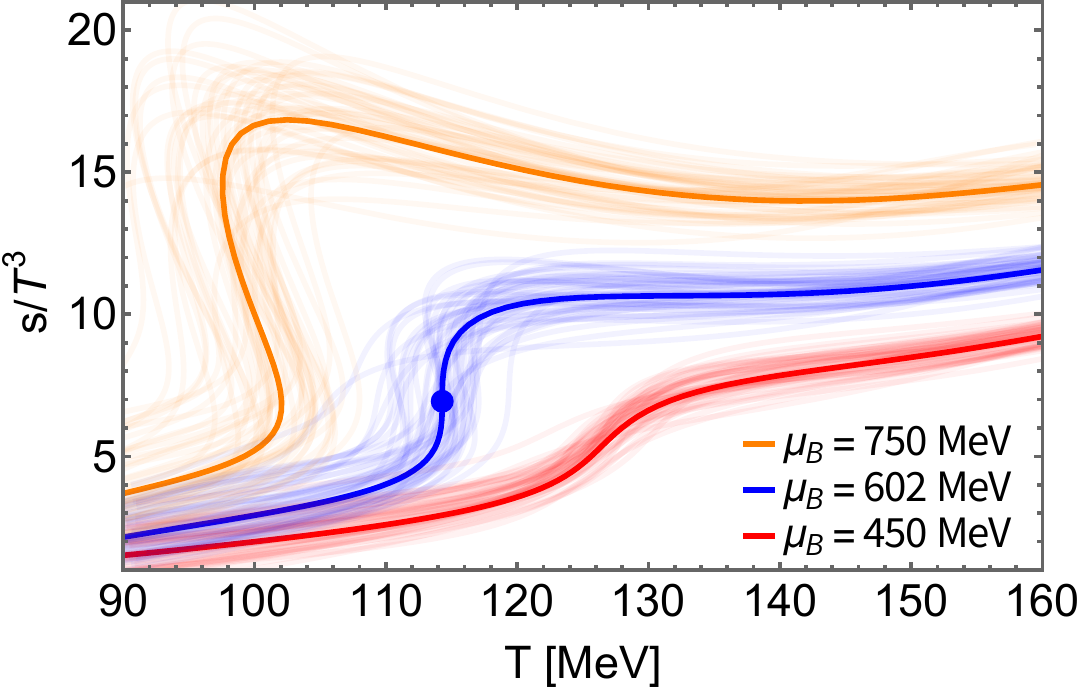}
    \caption{
    \small  
    Left panel: Results  for slices of constant $\mu_B/T$, with curves rescaled by different factors to prevent overlap and compared with available lattice QCD calctulations from Ref.\cite{Borsanyi:2021PRL}.
    Right panel: Results at fixed baryon chemical potentials: $\mu_B = 450$ MeV (red, crossover), $\mu_B = 602$ MeV (blue, critical), and $\mu_B = 750$ MeV (orange, mixed phase). The blue circle marks the location of the critical point.
    In both panels, solid lines denote the mean lattice QCD parametrization, while shaded (translucent) bands represent the propagated uncertainty from Monte Carlo sampling.}
    \label{fig:sT3}
    \vspace{-0.5cm}
\end{figure*}

Now that the methodology of the expansion is set, we apply it to the QCD equation of state employing lattice QCD results. For this, we parameterize $\chi_2^B$ and $s$ at $\mu_B=0$ from the lattice and compute $\alpha_2(T_0)$ using Eq.~\eqref{eq:C2}.  
After this, as we now have access to finite $\mu_B$ through the expansion Eq. \eqref{eq:TsO2}, we compare the scaled entropy as a function of temperature at constant $\mu_B/T$ slices with the already available state of the art lattice QCD calculations in Ref. \cite{Borsanyi:2021PRL} up to $\mu_B/T = 3.5$. 
We observe excellent agreement up to this value of chemical potential as seen in the left pane of Fig. \ref{fig:sT3}. 

The right panel of Fig.~\ref{fig:sT3} shows the behavior of $s/T^3$ when the expansion is pushed to larger chemical potentials, $\mu_B = 450$~MeV~(red), 602~MeV~(blue), and 750~MeV~(orange). 
One can see the development of a distinct S-shaped, multi-valued behavior of the entropy density as $\mu_B$ is increased to $\mu_B \gtrsim 602$~MeV -- a hallmark feature of a first-order phase transition.
Using Eq. \eqref{eq:spinodal} and a parameterized QCD input from the lattice, we obtain $T_{0,c} = 140.9 \pm 2.0$~MeV, and the CP at order $\mu_B^2$ at
\begin{equation}
(T_c,\mu_{B,c}) = (114.3 \pm 6.9, 602.1 \pm 62.1)~\text{MeV}
\end{equation}
shown as the solid blue circle in the figure. 
Error propagation was performed through Monte Carlo sampling (more details on the analysis can be found in the supplemental material of Ref. \cite{Shah:2024img}).

\begin{figure*}[!b]
\centering
\includegraphics[width=7cm,clip]{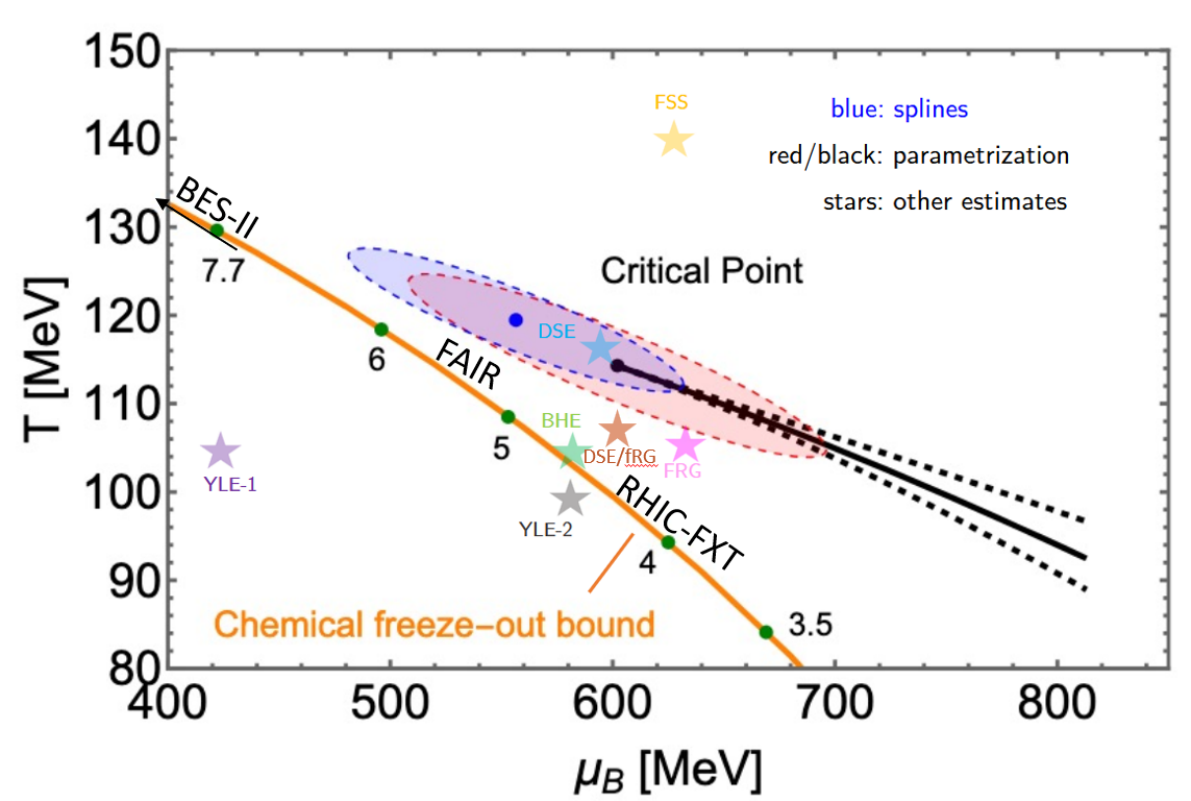}
 \caption{
The location of the QCD critical point extracted using contours of constant entropy density extrapolated from lattice QCD results using either parametric forms~(black point) or smoothing splines~(blue point) for  $s(T)$ and $\chi_2^B(T)$. 
The red and blue dashed ellipses denote the 68 \% confidence regions reflecting uncertainties from the lattice input for the former and latter methods, respectively.
The solid and dashed black curves represent the coexistence line and spinodal boundaries, respectively. The orange line shows the lower bound in temperature for the critical point
based on heavy-ion freeze-out conditions translated to $\mu_Q = \mu_S = 0$ conditions \cite{Lysenko:2024cfoc}. Green points denote chemical freeze-out parameters at various collision energies, while colored stars represent mean CP estimates from other theoretical \cite{Clarke:2024padetwo, Basar:2024qcd, Hippert:2024bhe, Fu:2020frg, Gunkel:2021dse, Gao:2021frg} and data-driven~\cite{Sorensen:2024fss} approaches.}
\label{CP-plot}       
\end{figure*}

The red ellipse in Fig.~\ref{CP-plot} shows the 68\% confidence region for the CP location, obtained using parameterized lattice QCD input, where estimates for the first-order and spinodal lines are also shown.
The uncertainties correspond exclusively to the Gaussian error propagation of the lattice QCD input and do not incorporate the truncation error. To assess systematic errors due to the parametrization of the lattice QCD results, we have repeated our analysis using instead a smoothed spline interpolation of the results for $s(T)$ and $\chi_2^B(T)$. Calculating $\alpha_2(T)$ from these splines, we find a CP at $(T_c,\mu_{B,c})=(119.5 \pm 5.4,556.5 \pm 49.8)$~MeV. Using Monte Carlo sampling to propagate errors leads to the blue ellipse, representing the 68\% confidence region for the CP location. 

\section{Conclusions}
We presented estimates for the QCD critical point based on leading-order extrapolation of constant-$s$ contours from $\mu_B = 0$ using lattice QCD data. 
The consistency across different parameterizations of the lattice results, which agree within one sigma, highlights the robustness of our approach. 
If the extrapolation holds, our results suggest that heavy-ion collisions at $\sqrt{s_{NN}} \approx$ $4-6$ GeV probe the region closest to the critical point. 
The current analysis is limited to leading-order $\mu_B^2$ expansion. Therefore, future studies incorporating for instance higher-order terms, analytic continuation from imaginary $\mu_B$  \cite{Borsanyi:2025dyp} , or even direct simulations at non-zero $\mu_B$, will be essential for refining the observed constraints.

\section*{Acknowledgments}
This material is based upon work supported by the National Science Foundation under grants No. PHY-2208724, PHY-2116686 and PHY-2514763 and within the framework
of the MUSES collaboration, under grant number No. OAC-2103680. This material is also based upon work supported by the U.S. Department of Energy, Office of Science, Office of Nuclear Physics, under Award Numbers DE-SC0022023 and DE-SC0026065, and by the National Aeronautics and Space Agency (NASA)  under Award Number 80NSSC24K0767. 
M.H. was supported in part by the Universidade Estadual do Rio de Janeiro through the Programa de Apoio à Docência (PAPD), and by the Brazilian National Council for Scientific and Technological Development (CNPq) under process No. 313638/2025-0. J.N. is partly supported by the U.S. Department of Energy, Office of Science, Office for Nuclear Physics under Award No. DE-SC0023861. 

\end{document}